\documentclass[twocolumn,aps,prd,amsmath,amssymb]{revtex4}
\bibpunct{}{}{,}{s}{}{,}
\usepackage{graphicx}   
\begin{document}

\title{Dynamics of black holes}

\author{Sean A. Hayward}
\affiliation{Center for Astrophysics, Shanghai Normal University, 100 Guilin 
Road, Shanghai 200234, China} 

\begin{abstract}
This is a review of current theory of black-hole dynamics, concentrating on the 
framework in terms of trapping horizons. Summaries are given of the history, 
the classical theory of black holes, the defining ideas of dynamical black 
holes, the basic laws, conservation laws for energy and angular momentum, other 
physical quantities and the limit of local equilibrium. Some new material 
concerns how processes such as black-hole evaporation and coalescence might be 
described by a single trapping horizon which manifests temporally as separate 
horizons. 

\end{abstract}

\maketitle

\section{Introduction}
Black holes are now generally regarded as astrophysical realities, which are 
expected to be major sources of gravitational radiation, prompting extensive 
studies of dynamical, strong-field processes such as binary mergers. The 
textbook theory of black holes, however, mostly concerns stationary black holes 
or physically unlocatable event horizons \cite{BCH,HE,MTW,Wal}. In recent 
years, a new paradigm for dynamical black holes has been developed in terms of 
trapping horizons, generated from surfaces where light is momentarily caught by 
the gravitational field. They locate the black hole in a practical, local way. 

The original idea due to Penrose \cite{Pen} is that of a trapped surface, which 
compose the interior of a Schwarzschild black hole. Such surfaces played a key 
role in the singularity theorems of Penrose and Hawking \cite{Pen,Haw,HP}. The 
boundary of the black hole is composed of marginally trapped surfaces, here 
called marginal surfaces. They are also commonly called apparent horizons, 
though the textbook definition of the latter differs \cite{HE,Wal}. By itself, 
this does not capture the idea of a black hole, as such surfaces also exist in 
cosmological models. Presumably for this reason, such quasi-local ideas were 
not much developed at the time. Instead history took a different turn, with the 
introduction of global methods and the near-universal acceptance of an event 
horizon as the definition of black hole. 

The author's interest in this area began with a rather simple observation: that 
it is outgoing (rather than ingoing) wavefronts which are marginally trapped, 
that ingoing wavefronts are converging, and that the trapped surfaces are 
inside (rather than outside). This led to a definition of a black hole by a 
future (rather than past) outer (rather than inner) trapping horizon 
\cite{bhd,bhs}. Here trapping horizon is used to mean any hypersurface foliated 
by marginal surfaces, a change from the original definition. 

Remarkably, this refinement sufficed to derive some key expected properties of 
black holes, assuming positive-energy conditions: the marginal surfaces have 
spherical topology; the horizon is spatial or null, and null only when a 
certain energy density vanishes, so that the horizon is one-way traversable; 
and the area $A$ is constant when null and increasing when spatial. Physically, 
if something falls into a black hole, the horizon moves out and the area 
increases. 

In the case of spherical symmetry, a rather complete picture emerged 
\cite{sph,1st,ine}. There is a standard definition \cite{MS} of active 
gravitational mass $M$. It satisfies an energy conservation law akin to the 
first law of thermodynamics, with energy-supply and work terms, reducing to the 
Bondi energy equation at null infinity. There is also a natural definition 
\cite{1st} of surface gravity $\kappa$, taking a quasi-Newtonian form, just 
$M/r^2$ in vacuo, in units $G=1$, where $A=4\pi r^2$. It has various desirable 
properties, including that $\kappa>0$, $\kappa=0$ or $\kappa<0$ on outer, 
degenerate or inner trapping horizons respectively. Then projecting the energy 
conservation law along a trapping horizon, denoted by a prime, yields 
$M'\cong\kappa A'/8\pi$ plus work term, where $\cong$ denotes evaluation on a 
trapping horizon. This is a dynamic version of the so-called first law of 
black-hole mechanics \cite{BCH}. 

Versions of these definitions and results also exist in cylindrical symmetry 
\cite{cyl}, which is the simplest situation allowing gravitational radiation, 
in plane symmetry \cite{pla} and in a quasi-spherical approximation 
\cite{qs,SH,gwbh,gwe}, which allows the interaction of roughly spherical black 
holes with gravitational radiation. In all these cases, there is an effective 
energy tensor $\Theta_{\alpha\beta}$ for the gravitational radiation, entering 
the energy conservation law additively with the matter energy tensor 
$T_{\alpha\beta}$. 

Meanwhile, Ashtekar and others developed a theory of isolated horizons 
\cite{ABF1,ABF2,AFK,ABD,ABL1,ABL2,Boo1,GJ1,DKSS}, which are null trapping 
horizons with a hierarchy of additional conditions, each intended to describe a 
black hole in local equilibrium to some degree, generalizing Killing horizons. 
In particular, they defined angular momentum $J$, entering a first law in the 
expected way. 

A breakthrough came with the work of Ashtekar \& Krishnan \cite{AK1,AK2,AK3} on 
what they called dynamical horizons, which are spatial future trapping 
horizons. They derived flux laws for both energy and angular momentum. In the 
versions subsequently developed by the author \cite{bhd2,bhd3,bhd4,bhd5}, they 
hold for any trapping horizon and take the form of conservation laws 
\begin{eqnarray}
L_\xi
M&\cong&\oint_S{*}(T_{\alpha\beta}+\Theta_{\alpha\beta})k^\alpha\tau^\beta\label{enc}\\
L_\xi
J&\cong&-\oint_S{*}(T_{\alpha\beta}+\Theta_{\alpha\beta})\psi^\alpha\tau^\beta\label{amc}
\end{eqnarray}
where $S$ is a marginal surface, ${*}1$ its area form, $L$ is the Lie 
derivative, $\xi$ is a generating vector of the trapping horizon, $\tau$ its 
normal dual \cite{gr,amf,BHMS}, and $k$ and $\psi$ are certain vectors which 
play the role normally played by Killing vectors. Here $\Theta_{\alpha\beta}$ 
is again interpreted as an effective energy tensor for gravitational radiation. 
Apart from the inclusion of $\Theta_{\alpha\beta}$, these have the same form as 
corresponding flat-space conservation laws in surface-integral form. 

Some subtleties remained. Firstly, in what one might call the bottom-up 
approach of Ashtekar and co-authors, the formalisms for isolated and dynamical 
horizons were quite different, and it was not clear how to relate them. Some 
progress was made by Booth \& Fairhurst for slowly evolving horizons 
\cite{BF1,BF2,BF3}. Conversely, in the author's top-down approach, whereby 
local equilibrium is given simply by the trapping horizon being null, there is 
a degeneracy in taking the null limit. Remarkably, consideration of angular 
momentum closed this gap by suggesting a natural way to fix the degeneracy 
\cite{bhd4,bhd5}. The resulting gauge-fixed null trapping horizon is then 
equivalent, modulo further gauge-fixing, to a weakly isolated horizon 
\cite{ABL1,ABL2,Boo1,GJ1}. Thus the top-down and bottom-up approaches have 
converged on the same definition of a black hole in local equilibrium. 

For this reason, combined with the existence of the reviews of Ashtekar \& 
Krishnan \cite{AK3}, Booth \cite{Boo2}, Krishnan \cite{Kri} and Gourgoulhon \& 
Jaramillo \cite{GJ2}, this review will concentrate on the paradigm in terms of 
trapping horizons, without attempting to make detailed comparisons with the 
other approaches. Indeed, it is intended to be largely complementary to those 
reviews, apart from inevitable overlap on core issues. No attempt will be made 
at a comprehensive list of references, since the area and related areas are 
rapidly evolving. With a broad audience in mind, detailed calculations and 
proofs will be deferred to references. General Relativity, including the 
Einstein equation, will be assumed throughout, though the essential ideas can 
be generalized to other metric theories or dimensions. 
 
\section{Classical theory of black holes}

What one might nowadays call Newtonian black holes were described long ago by 
Mitchell and Laplace \cite{HE}, as objects for which the escape speed 
$\sqrt{2GM/r}$ is greater than the speed $c$ of light. Any light leaving such 
an object would be trapped by the gravitational field and fall back, rendering 
it invisible. Curiously, the Newtonian relation $r<2GM/c^2$ will survive quite 
generally in a sense to be revealed, with units $c=1=G$ henceforth. 

Almost immediately after Einstein formulated General Relativity \cite{Ein} in 
1915, Schwarzschild \cite{Sch} derived the solution for the gravitational field 
of a point mass $M$. Charge $Q$ was soon added by Reissner \cite{Rei} and 
Nordstr\"om \cite{Nor}. However, the geometry of the solutions was not 
understood for decades. Einstein \& Rosen \cite{ER} deciphered the ($Q=0$) 
spatial geometry in 1935, with a wormhole connecting two asymptotically flat 
spaces, but it was not until around 1960 that the maximally extended space-time 
geometry was understood, with work of Wheeler, Kruskal \cite{Kru} and Fronsdal 
\cite{Fro}. The space-time diagram clearly shows a region $r<2M$ from which 
light cannot escape. It had already been shown by Oppenheimer \& Snyder 
\cite{OS} in 1939 that gravitational collapse could indeed produce such a 
region. According to relativistic causality, nothing else can escape either. 

Angular momentum $J$ was added by Kerr \cite{Ker} and (for $Q\not=0$) Newman et 
al.\ \cite{New} and finally around 1968, the term ``black hole'' entered the 
lexicon, attributed to Wheeler \cite{Whe}. A few years of rapid progress 
followed, culminating in the classical paradigm. Black holes were defined in 
general by event horizons as defined by Penrose \cite{Pen2}, which satisfy the 
area-increase theorem of Hawking \cite{Haw2}, $A'\ge0$. This became known as a 
second law, part of the four laws of black-hole mechanics formulated by 
Bardeen, Carter \& Hawking \cite{BCH}, which seemed analogous to the four laws 
of thermodynamics. The zeroth law is that surface gravity $\kappa$ is constant 
on stationary black holes. The first law is 
\begin{equation}
\delta E=\kappa\delta A/8\pi+\Omega\delta J+\Phi\delta Q \label{first0}
\end{equation} 
for perturbations of stationary black holes, where $E$ is the ADM energy at 
spatial infinity, $\Omega$ is the angular speed and $\Phi$ is the electric 
potential. The third law is $\kappa\not\to0$, by perturbations of stationary 
black holes. The discovery of quantum black-hole radiance by Hawking 
\cite{Haw3}, with temperature $\kappa/2\pi$ in units $k=1$, led to much 
research on black holes as a key area to generate and test ideas concerning the 
interface of gravity, quantum theory and thermodynamics. 

\begin{figure}
\includegraphics[height=6cm]{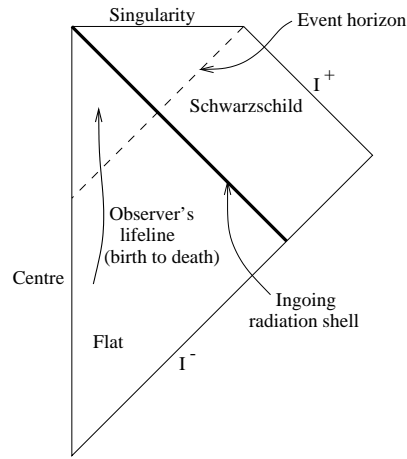}
\caption{Example of the unknowable nature of event horizons:
Penrose diagram of a spherically symmetric space-time
which is initially flat but contains an ingoing radiation shell,
forming a Schwarzschild black hole.
An observer in the flat region crosses the event horizon but feels nothing.}
\label{event}
\end{figure}

Once a theoretical curiosity, observational evidence for black holes has by now 
accumulated past the point where they are part of mainstream astrophysics, both 
as stellar-mass black holes produced as supernova remnants, and as supermassive 
black holes mysteriously found at the cores of most galaxies, powering active 
galactic nuclei. Gravitational-wave detectors are expected to be able to 
observe such processes as binary inspiral and merger of black holes or neutron 
stars, prompting much research in numerically computing waveforms. 

In such dynamical situations, the classical paradigm is of little help. 
Firstly, event horizons are defined as the boundary between where light can or 
cannot escape to infinity, which means not only infinite distance but 
infinitely far in the future. Therefore one cannot locate it without knowing 
future events, indeed the entire future history of the universe, which is 
beyond mere mortals. It is easy to give examples \cite{mg9} where an event 
horizon exists in a region of flat space-time (Fig.~\ref{event}). Then an 
observer crossing the event horizon cannot detect it in any way and has never 
experienced a gravitational field. 

In practice, it is normally much easier to locate marginal surfaces in 
numerical simulations. Note also that three of the above laws are more 
accurately described as black-hole statics, since they concern stationary black 
holes, specifically Killing horizons. So the question arises: are there 
corresponding laws of black-hole dynamics? 
 
\section{Dynamical black holes}
The basic idea is easily explained. Imagine enclosing a star with a roughly 
spherical spatial surface at some moment of time, and detonating a flash of 
light simultaneously at each point of the surface. Two wavefronts form, one 
ingoing and one outgoing. Normally one expects the outgoing wavefront to have 
increasing area and the ingoing wavefront to have decreasing area. This is 
measured at each point by the expansions $\theta_\pm$ of light rays in the 
wavefronts: $\theta_+>0$ for the outgoing wavefront and $\theta_-<0$ for the 
ingoing wavefront. However, the gravitational field of the star tends to drag 
things toward it, including light. Thus the outgoing wavefront does not expand 
as much as if the star were not present. The effect increases closer to the 
star and for larger mass, as well known from Newton's law of gravitation. For 
large enough mass, for a surface close enough, it may happen that the outgoing 
wavefront has decreasing area. This is a black hole: all light and therefore 
matter are confined inside a shrinking area. 

The idea then is that the boundary of the black hole at a given time is a 
marginal surface, meaning that one wavefront has instantaneously parallel light 
rays, in this case $\theta_+=0$. To characterize a black hole, it is also 
important that the ingoing wavefront is converging, $\theta_-<0$, and that 
$\theta_+$ is decreasing in the ingoing direction. 

This is formulated geometrically as follows. A spatial surface $S$ in 
space-time has two unique future-pointing null normal directions, along null 
normal vectors $l_\pm$: 
\begin{equation} 
g(l_\pm,l_\pm)=0,\quad\bot l_\pm=0
\end{equation} 
where $g$ is the space-time metric and $\bot$ denotes projection onto $S$. The 
null expansions are
\begin{equation}
\theta_\pm=L_\pm\log{*}1
\end{equation} 
where $L_\pm=L_{l_\pm}$.

Then $S$ is said to be {\em trapped} if $\theta_+\theta_->0$, {\em marginal} if 
$\theta_+=0$ or $\theta_-=0$, and {\em untrapped} (or mean convex) if 
$\theta_+\theta_-<0$. In terms of the expansion vector or mean-curvature vector 
\begin{equation}
H=g^{-1}(d\log{*}1)
\end{equation} 
this is equivalent to $H$ being temporal, null or spatial, respectively. One 
finds 
\begin{equation}
H=-e^f(\theta_-l_++\theta_+l_-)
\end{equation} 
where $f$ is a normalization function, 
\begin{equation} 
e^{-f}=-g(l_+,l_-). \label{f}
\end{equation} 
For any normal vector $\eta$, $\bot\eta=0$, the expansion 
$\theta_\eta=L_\eta\log{*}1$ is given by $\theta_\eta=g(H,\eta)$. 

Untrapped surfaces have a local spatial orientation: an {\em achronal} (meaning 
spatial or null) normal vector $\eta$ is {\em outward} or {\em inward} if 
$\theta_\eta>0$ or $\theta_\eta<0$ respectively. Locally one can conventionally 
fix $\theta_+>0$, $\theta_-<0$ in an untrapped region; then $l_+$ is outward 
and $l_-$ is inward. Conversely, trapped surfaces have a local causal 
orientation: if $H$ is future or past {\em causal} (meaning temporal or null), 
the surface is {\em future} or {\em past} trapped respectively. Then future or 
past trapped surfaces have $\theta_\pm<0$ or $\theta_\pm>0$ respectively. 
Marginal surfaces have both orientations, if the other null expansion has fixed 
non-zero sign. 

A {\em trapping horizon} is a hypersurface foliated by marginal surfaces. It is 
said to be {\em outer} or {\em inner} if $L_-\theta_+<0$ or $L_-\theta_+>0$ 
respectively, for the case $\theta_+=0$, where the null normals $l_\pm$ are 
extended off the horizon to generate ingoing and outgoing wavefronts from the 
marginal surfaces, i.e.\ two families of null hypersurfaces labelled by 
$x^\pm$, intersecting in the marginal surfaces, such that 
$l_A(dx^B)=\delta_A^B$. This is locally unique unless the trapping horizon is 
itself null. Equivalently, the horizon is outer or inner if 
$\nabla\cdot(H-H^*)>0$ or $\nabla\cdot(H-H^*)<0$ respectively, where 
$\eta^*=\eta^+l_+-\eta^-l_-$ is the normal dual vector \cite{gr,amf,BHMS} to a 
normal vector $\eta=\eta^+l_++\eta^-l_-$: 
\begin{equation}
\bot\eta^*=0,\quad g(\eta^*,\eta)=0,\quad g(\eta^*,\eta^*)=-g(\eta,\eta).
\end{equation} 
This is a vector version of the normal Hodge dual for 1-forms. 

Then a future (respectively past) outer trapping horizon provides a local 
definition of a generic black (respectively white) hole \cite{bhd}. More 
precisely, the idea is that a non-degenerate black hole exists only if such a 
horizon exists. As to the converse, see the Concluding Remarks. Note that outer 
versus inner has been defined with respect to the ingoing null direction, as 
this is invariant, rather than in some ingoing spatial direction. Also, 
demanding a strict sign for $L_-\theta_+$ has excluded extremal black holes, 
for which $L_-\theta_+$ vanishes. This is for simplicity only, as they can be 
treated as special cases. Relaxing the definition to $L_-\theta_+\le0$ would 
allow cases which are not black holes. 

\section{Basic laws}
Summarized here are some basic properties of trapping horizons \cite{bhd}. 
Where not obvious, the proofs involve either (for $\theta_+=0$) the $T_{++}$ 
component of the Einstein equation, where the null energy condition (NEC) is 
cited, or the $T_{+-}$ component, where the dominant energy condition (DEC) is 
cited. One introduces a normal generating vector $\xi$ of the marginal surfaces 
in the horizon, which therefore satisfies $L_\xi\theta_+=0$ on the horizon. The 
area of the marginal surfaces, if compact, is 
\begin{equation} 
A=\oint_S{*}1. 
\end{equation} 

{\em Trapping}: for a future/past, outer/inner trapping horizon, there are 
trapped surfaces to one side and untrapped surfaces to the other side. For a 
future outer trapping horizon, this reflects the defining idea that outgoing 
light rays are momentarily parallel, $\theta_+=0$, diverging just outside, 
$\theta_+>0$, and converging just inside, $\theta_+<0$, while ingoing light 
rays are converging, $\theta_-<0$. 

{\em Signature}: assuming NEC, an outer or inner trapping horizon is achronal 
or causal respectively, and null if and only if the effective ingoing energy 
density $T_{++}+\Theta_{++}$ vanishes, where the explicit expressions for 
$\Theta_{\alpha\beta}$ are given later (\ref{Theta}). In particular, this means 
that {\em black-hole horizons are one-way traversable}: one can fall into a 
black hole but not escape, at least through the outer horizon. 

{\em Area}: assuming NEC, future outer or past inner trapping horizons have 
non-decreasing area form, $\theta_\xi\ge0$, and therefore (if compact) 
non-decreasing area, $L_\xi A\ge0$, instantaneously constant ($\theta_\xi=0$) 
if and only if the horizon is null. For past outer or future inner trapping 
horizons, all signs reverse. Here the orientation of $\xi$ is such that, in the 
null limit, it is future-null, which means that it is future causal for inner 
horizons and outward achronal for outer horizons. In particular, this means 
that {\em black holes grow} if they absorb any matter or gravitational 
radiation, and otherwise remain the same size. 

{\em Topology}: assuming DEC, a future/past outer trapping horizon has marginal 
surfaces of spherical topology (if compact). The proof uses the Gauss-Bonnet 
and Gauss divergence theorems. Thus {\em realistic black holes are 
topologically spherical}. If degenerate horizons are considered, then toroidal 
topology is just allowed, but highly non-generic, in particular Gaussian flat, 
and so presumably unstable. 

{\em Area limit}: assuming DEC and a positive cosmological constant $\Lambda$, 
outer trapping horizons satisfy\cite{HSN} $A\le4\pi/\Lambda$. Thus black holes 
are smaller than the cosmological horizon scale, corresponding to an area 
$12\pi/\Lambda$. 

Pedagogically, this would be the place to give details in spherical symmetry 
\cite{sph,1st,ine}, cylindrical symmetry \cite{cyl}, plane symmetry \cite{pla} 
and the quasi-spherical approximation \cite{qs,SH,gwbh,gwe}, but they are 
omitted here due to limitations of space, apart from as explained in the 
Introduction. 

\section{Conservation of energy}
The simplest generalization of the Schwarzschild relation $1-2M/r=g^{rr}$ to a 
general surface is the Hawking mass \cite{Haw4}
\begin{eqnarray} 
M&=&\frac r2\left(1-\frac1{16\pi}\oint_S{*}g(H,H)\right)\\
&=&\frac r2\left(1+\frac1{8\pi}\oint_S{*}e^f\theta_+\theta_-\right)\nonumber
\end{eqnarray} 
where 
\begin{equation} 
r=\sqrt{A/4\pi}
\end{equation} 
is the area radius. It has various useful properties. Large spheres: in an 
asymptotically flat space-time, $M$ tends to the Bondi or ADM energy at null or 
spatial infinity, respectively. Small spheres: $M/\hbox{volume}\to$ density at 
a regular centre. Trapping: a surface is trapped, marginal or untrapped if 
$r<2M$, $r=2M$ or $r>2M$ respectively. 

In particular, this means that $M$ is the {\em irreducible mass} of a future 
outer trapping horizon, $L_\xi M\ge0$, assuming NEC. This follows directly from 
the area law, since $A\cong4\pi(2M)^2$, where $\cong$ henceforth denotes 
evaluation on a marginal surface. Recall the irreducible mass for stationary 
black holes \cite{Chr}: $M\cong(\frac12m(m+(m^2-a^2)^{1/2})^{1/2}$ for Kerr 
black holes, the mass which remains even if rotational energy is removed by the 
Penrose process. The original concept arose due to quasi-stationary arguments, 
but here it is exact and gives a physical meaning to $M$. 

\begin{figure}
\includegraphics[height=4cm]{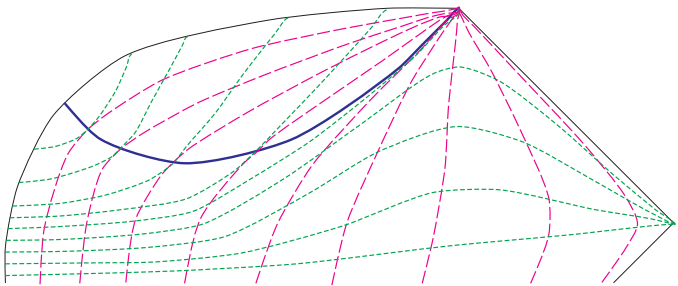}
\caption{Penrose diagram of gravitational collapse to a black hole, satisfying 
cosmic censorship. When a trapping horizon (blue bold line) forms, the flow 
lines of $k$ (magenta long dashes) and $g^{-1}(dr)$ (green short dashes) switch 
over, both being null at the horizon.} \label{collapse} 
\end{figure}

The simplest generalization of the Schwarzschild stationary Killing vector
$k=\partial_t$ is the canonical time vector \cite{bhd2}
\begin{equation} 
k=(g^{-1}(dr))^*=e^f(L_+r\,l_--L_-r\,l_+)
\end{equation} 
or equivalently
\begin{equation} 
\bot k=0,\quad k\cdot dr=0,\quad g(k,k)=-g^{-1}(dr,dr).
\end{equation} 
In particular, $g(k,k)\cong0$ and $k\cong\pm g^{-1}(dr)$ on a trapping horizon 
$\theta_\pm\cong0$. So trapping horizons are characterized by $k$ being null, 
just as Killing horizons are characterized by a stationary Killing vector being 
null (Fig.\ref{collapse}). In spherical symmetry, $k$ reduces to the Kodama 
vector \cite{Kod}, which \cite{sph} is a Noether current with Noether charge 
$M$, and is related \cite{1st} to $\kappa$ by 
$k^\beta\nabla_{[\beta}k_{\alpha]}\cong\pm\kappa k_\alpha$ on a trapping 
horizon $\theta_\pm\cong0$, analogously to the usual definition of surface 
gravity for stationary black holes. 

\begin{figure}
\includegraphics[height=15mm]{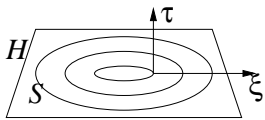}
\caption{A non-null hypersurface $H$ foliated by spatial surfaces $S$, with
generating vector $\xi$ and its normal dual $\tau=\xi^*$.} \label{xitau}
\end{figure}

Lastly, one needs the dual normal vector to the trapping horizon: 
\begin{equation} 
\tau=\xi^*=\xi^+l_+-\xi^-l_-
\end{equation} 
which can be locally chosen to be future-pointing for outward-pointing $\xi$. 
For a spatial trapping horizon, $\xi$ is spatial and $\tau$ is temporal, while 
$\tau\to\xi$ as a trapping horizon becomes null (Fig.\ref{xitau}). 

With these ingredients, conservation of energy takes the surface-integral form 
\cite{bhd2,bhd3} 
\begin{equation} 
L_\xi M\cong\oint_S{*}(T_{AB}+\Theta_{AB})k^A\tau^B
\end{equation} 
where the components $\Theta_{AB}$ are given below. The proof is a calculation 
using the $T_{++}$ and $T_{+-}$ components of the Einstein equation and  the 
Gauss-Bonnet and Gauss divergence theorems. It may be noted that evaluation on 
a trapping horizon yields remarkable cancellations in what would otherwise be 
lengthy expressions on the right-hand side. The conservation law can 
equivalently be written in volume-integral form 
\begin{equation} 
[M]\cong\int_H{*}(T_{AB}+\Theta_{AB})k^A\tau^B\wedge dx
\end{equation} 
where $x$ labels the marginal surfaces, $\xi=\partial_x$, which expresses the 
change $[M]$ in $M$ between two marginal surfaces in the horizon $H$. For a 
spatial trapping horizon with unit normal $\hat\tau=\tau/\sqrt{g_{xx}}$ and 
proper volume element $\hat{*}1={*}\sqrt{g_{xx}}\wedge dx$, there is the 
proper-volume form 
\begin{equation} 
[M]\cong\int_H\hat{*}(T_{AB}+\Theta_{AB})k^A\hat\tau^B
\end{equation} 
which has the same form as the usual expression for energy, with $k$ replacing 
a stationary Killing vector. However $\hat\tau$ is ill defined and 
$\hat{*}1\to0$ if the trapping horizon becomes null, $g_{xx}\to0$, which is the 
physically important limit where a growing black hole ceases to grow, so the 
above surface-integral form is preferred. 

To spell out the components $\Theta_{AB}$, introduce the transverse metric 
$h_{ab}$, i.e.\ the induced metric of $S$, the null shears 
\begin{equation} 
\sigma_{\pm ab}=h_a^\gamma h_b^\delta L_\pm h_{\gamma\delta}-\theta_\pm h_{ab}
\end{equation} 
and the normal fundamental forms 
\begin{equation} 
\zeta_{\pm a}=e^fh_a^\gamma(l_{\pm\beta}\nabla_\gamma l_\mp^\beta).
\end{equation} 
Note here that indices $\alpha$, $\beta\ldots$ are general, $A$, $B\ldots$ 
normal, and $a$, $b\ldots$ transverse. Then the symmetric bilinear forms 
$\sigma_\pm$ are transverse, $\sigma_\pm=\bot\sigma_\pm$, and traceless, 
$h^{ab}\sigma_{\pm ab}=0$, while $\zeta_\pm$ are transverse 1-forms,  
$\zeta_\pm=\bot\zeta_\pm$. Then the components of $\Theta_{\alpha\beta}$ with 
respect to $\l_\pm$ are 
\begin{equation} 
\Theta_{\pm\pm}=||\sigma_\pm||^2/32\pi,\quad 
\Theta_{\pm\mp}=e^{-f}|\zeta_\pm|^2/8\pi\label{Theta}
\end{equation} 
where $|\zeta|^2=h^{ab}\zeta_a\zeta_b\ge0$, 
$||\sigma||^2=h^{ab}h^{cd}\sigma_{ac}\sigma_{bd}\ge0$ are transverse norms. 
Here the signs indicate that $\Theta_{\alpha\beta}$ satisfies DEC, so that {\em 
gravitational radiation carries positive energy}. 

The components may be interpreted as gravitational energy densities, by 
geodesic deviation of test particles \cite{Sze}: $\Theta_{++}$ is the ingoing 
transverse mode, reducing to the Bondi energy density at past null infinity; 
$\Theta_{--}$ is the outgoing transverse mode, reducing to the Bondi energy 
density at future null infinity; $\Theta_{-+}$ is the ingoing longitudinal 
mode, with $r^2\Theta_{-+}\to0$ at null infinity; $\Theta_{+-}$ is the outgoing 
longitudinal mode, with $r^2\Theta_{+-}\to0$ at null infinity. The 
$\Theta_{\pm\pm}$ components also recover expressions for energy density of 
gravitational radiation, with $\Theta_{\pm\mp}$ vanishing, in the 
high-frequency linearized approximation \cite{MTW}, in cylindrical symmetry 
\cite{cyl}, plane symmetry \cite{pla} and in the quasi-spherical approximation 
\cite{qs,SH,gwbh,gwe}. 

Physically, the energy conservation law expresses the increase in irreducible 
black-hole mass $M$ in terms of the energy densities of the infalling matter 
and gravitational radiation. Since $A\cong4\pi(2M)^2$, it also describes {\em 
how a black hole grows}. 

\section{Conservation of angular momentum}
The standard definition of angular momentum for an axial Killing vector $\psi$ 
and at spatial infinity is the Komar integral \cite{Kom}
\begin{equation} 
J[\psi]=-\frac1{16\pi}\oint_S{*}\epsilon_{\alpha\beta}\nabla^\alpha\psi^\beta
\end{equation} 
where $\epsilon_{AB}$ is the binormal. For a general transverse vector $\psi$, 
$\bot\psi=\psi$, it can be rewritten as \cite{bhd4,bhd5} 
\begin{equation} 
J[\psi]=\frac1{8\pi}\oint_S{*}\psi^a\omega_a
\end{equation} 
where the twist \cite{dne}
\begin{equation} 
\omega_a=\frac12e^fh_{a\beta}[l_-,l_+]^\beta
\end{equation} 
is a transverse 1-form, $\bot\omega=\omega$, measuring the non-integrability of 
the normal space. 

For the weak-field metric in spherical polar coordinates 
$(t,r,\vartheta,\varphi)$, $J[\partial_\varphi]$ recovers the standard 
definition of angular momentum \cite{MTW}. Also the precessional angular 
velocity $(\vec\Omega\cdot\hat r)\hat r-\frac13\vec\Omega$ of a gyroscope in 
the unit direction $\hat r$, due to the Lense-Thirring effect, is directly 
related to the twist by $\omega\sim\vec\Omega\times\hat r$. Thus the twist 
indeed encodes the twisting around of space-time due to a rotating mass. 

The twist is an invariant of a non-null foliated hypersurface $H$, so the twist 
expression for $J[\psi]$ is also an invariant of $H$. It coincides with the 
1-form used to define angular momentum for dynamical horizons by Ashtekar \& 
Krishnan \cite{AK1}, but not that used for isolated horizons by Ashtekar et 
al.\ \cite{ABD,ABL1}, which is $\omega+\frac12Df$, where $D$ is the covariant 
derivative of $h$. They will give compatible $J[\psi]$, by the Gauss divergence 
theorem, if the axial vector has vanishing transverse divergence, 
\begin{equation} 
D_a\psi^a\cong0.\label{div}
\end{equation} 
If there exist angular coordinates $(\vartheta,\varphi)$ on $S$, completing 
coordinates $(x,\vartheta,\varphi)$ on $H$, such that $\psi=\partial_\varphi$, 
then recalling that $\xi=\partial_x$ and that coordinate vectors commute, 
\begin{equation} 
L_\xi\psi\cong0\label{lie}
\end{equation} 
which was previously proposed as a natural way to propagate $\psi$ along $\xi$ 
by Gourgoulhon \cite{Gou}. There have been various suggestions for specifying 
$\psi$ more uniquely \cite{bhd4,bhd5,CW,Kor}. 

Then conservation of angular momentum takes the form \cite{bhd4,bhd5}
\begin{equation} 
L_\xi J\cong-\oint_S{*}(T_{aB}+\Theta_{aB})\psi^a\tau^B.
\end{equation} 
The proof is a calculation using the $T_{aB}$ components of the Einstein 
equation, and requires various cancellations due to evaluation on a trapping 
horizon and the conditions (\ref{div})--(\ref{lie}). Here 
\begin{equation} 
\Theta_{a\pm}=-{1\over{16\pi}}h^{cd}D_d\sigma_{\pm ac}
\end{equation} 
is the transverse-normal block of the effective energy tensor for gravitational 
radiation. Recalling the energy densities 
$\Theta_{\pm\pm}=||\sigma_\pm||^2/32\pi$ (\ref{Theta}), indicating that 
transverse gravitational radiation is encoded in null shear $\sigma_\pm$, it 
seems that {\em differential gravitational radiation has angular momentum 
density}. So this describes {\em how a black hole spins up or down}, due to 
infall of co-rotating or counter-rotating matter or gravitational radiation. 

Thus conservation of energy and angular momentum take a similar form 
(\ref{enc}), (\ref{amc}). Both take the same form as standard expressions in 
flat space-time for a stationary Killing vector $k$ and an axial Killing vector 
$\psi$, except for the inclusion of gravitational radiation in 
$\Theta_{\alpha\beta}$. They are the independent conservation laws expected for 
an astrophysical black hole, which defines its own centre-of-mass frame and its 
own axis of rotation. 

Higher source multipoles have been defined for isolated horizons by Ashtekar et 
al.\ \cite{AEPV} and for dynamical horizons by Schnetter et al.\ \cite{SKB}. A 
gauge-dependent definition of linear momentum has been proposed by Krishnan et 
al.\ \cite{KCZ}, which may be useful in studying the recoil or kick effect in 
asymmetric binary mergers. 

\section{Quasi-local conservation laws}
To compare with the classical paradigm, one needs to include charge $Q$, which 
is defined in terms of charge-current density $j$ as 
\begin{equation} 
[Q]=-\int_H{*}g(j,\tau)\wedge dx=-\int_H{\hat*}g(j,\hat\tau)
\end{equation} 
where the more usual second expression holds only for a spatial hypersurface 
$H$. As above, the surface-integral form is 
\begin{equation} 
L_\xi Q=-\oint_S{*}g(j,\tau).
\end{equation} 
The above conservation laws can be written in the same form 
\begin{equation} 
L_\xi M\cong-\oint_S{*}g(\bar\jmath,\tau),\quad L_\xi 
J\cong-\oint_S{*}g(\tilde\jmath,\tau)
\end{equation} 
by identifying current vectors
\begin{equation} 
\bar\jmath^B=-k_A(T^{AB}+\Theta^{AB}),\quad
\tilde\jmath^B=\psi_a(T^{aB}+\Theta^{aB}).
\end{equation} 
The standard physical interpretation of the conserved vectors is 
$\bar\jmath=({}$energy density, energy flux${})$, $\tilde\jmath=({}$angular 
momentum density, angular stress${})$, $j=({}$charge density, current 
density${})$. For spatial $\xi$: $\oint_S{*}g(\bar\jmath,\xi)={}$power;  
$\oint_S{*}g(\tilde\jmath,\xi)={}$torque; $\oint_S{*}g(j,\xi)={}$current; 
$-\oint_S{*}g(\bar\jmath,\tau)={}$energy gradient;
$-\oint_S{*}g(\tilde\jmath,\tau)={}$angular momentum gradient; 
$-\oint_S{*}g(j,\tau)={}$charge gradient. 

Charge conservation in local differential form is 
\begin{equation} 
\nabla_\alpha j^\alpha=0.
\end{equation} 
However, for energy and angular momentum, one has only quasi-local conservation 
laws holding on a trapping horizon: 
\begin{equation} 
\oint_S{*}\nabla_\alpha\bar\jmath^\alpha\cong\oint_S{*}\nabla_\alpha\tilde\jmath^\alpha\cong0.
\end{equation} 
This subtly confirms the view that energy and angular momentum in General 
Relativity cannot be localized \cite{MTW}, but might be quasi-localized as 
surface integrals, as long ago argued by Penrose \cite{Pen3}. The corresponding 
conservation laws have indeed been obtained in surface-integral but not local 
form. 

\section{State space} 
There are now three conserved quantities $(M,J,Q)$, forming a state space for 
dynamical black holes. Following various authors 
\cite{ABF1,ABF2,AFK,ABD,ABL1,ABL2,Boo1,GJ1,DKSS,AK1,AK2,AK3,BF1,BF2,BF3,Boo2,Kri}, 
related quantities may then be defined by formulas satisfied by Kerr-Newman 
black holes, specifically those for the ADM energy 
\begin{equation} 
E\cong\frac{\sqrt{((2M)^2+Q^2)^2+(2J)^2}}{4M} \label{e}
\end{equation}
the surface gravity
\begin{equation}
\kappa\cong\frac{(2M)^4-(2J)^2-Q^4}{2(2M)^3\sqrt{((2M)^2+Q^2)^2+(2J)^2}}
\label{sg}
\end{equation}
the angular speed
\begin{equation}
\Omega\cong\frac{J}{M\sqrt{((2M)^2+Q^2)^2+(2J)^2}} \label{as}
\end{equation}
and the electric potential
\begin{equation}
\Phi\cong\frac{((2M)^2+Q^2)Q}{2M\sqrt{((2M)^2+Q^2)^2+(2J)^2}}. \label{ep}
\end{equation}
These formulas may not be so familiar, since the classical theory uses 
$(E,J,Q)$ as parameters with $M$ dependent, but are easily obtained by 
inverting standard formulas. 

In the dynamical context, $E\ge M$ is not the ADM energy, but can be 
interpreted as the {\em effective energy} of the black hole. Expanding for 
$J\ll M^2$ and $Q\ll M$, $E\approx M$ to leading order, then to next order,
\begin{equation}
E\approx M+\textstyle{\frac12}I\Omega^2+\textstyle{\frac12}Q^2/r
\end{equation}
where $J=I\Omega$ defines the moment of inertia $I\cong 
M\sqrt{((2M)^2+Q^2)^2+(2J)^2}\cong Er^2$. Thus $E$ includes irreducible mass 
$M$, rotational kinetic energy $\approx\frac12I\Omega^2$ and electrostatic 
energy $\approx\frac12Q^2/r$, the latter being standard Newtonian expressions. 

The state-space formulas 
\begin{equation}
\kappa\cong8\pi\frac{\partial E}{\partial A}\cong{1\over{4M}}\frac{\partial
E}{\partial M}, \quad\Omega\cong\frac{\partial E}{\partial J},
\quad\Phi\cong\frac{\partial E}{\partial Q} \label{pd}
\end{equation}
then yield a dynamic version \cite{bhd4,bhd5} of the so-called first law of 
black-hole mechanics (\ref{first0}): 
\begin{equation}
L_\xi E\cong\frac{\kappa}{8\pi}L_\xi A+\Omega L_\xi J+\Phi L_\xi Q. \label{g}
\end{equation}
Here the state-space perturbations in the classical law for Killing horizons 
\cite{BCH}, or the versions for isolated horizons 
\cite{ABF1,ABF2,AFK,ABD,ABL1,ABL2,Boo1,GJ1}, have been replaced by derivatives 
along the trapping horizon, thereby promoting it to a dynamical law. 

\section{Equilibrium: null trapping horizons} 
When a growing black hole ceases to grow, the generically spatial trapping 
horizon becomes null. The central result here is that, assuming DEC, 
\begin{equation}
g(\bar\jmath,\tau)\cong g(\tilde\jmath,\tau)\cong g(j,\tau)\cong0
\end{equation}
on a null trapping horizon \cite{bhd4,bhd5}. Therefore the conserved quantities 
are actually preserved: 
\begin{equation}
L_\xi M\cong L_\xi J\cong L_\xi Q\cong0.
\end{equation} 
This indicates that {\em local equilibrium is attained when a trapping horizon 
becomes null}. 

It follows that 
\begin{equation}
L_\xi E\cong L_\xi\kappa\cong L_\xi\Omega\cong L_\xi\Phi\cong0.
\end{equation}
In particular, the surface gravity, which already satisfies $D\kappa\cong0$ by 
definition (\ref{sg}), is constant where a trapping horizon becomes null. This 
is a quite general zeroth law. The above result is stronger still, since it 
expresses complete local equilibrium, not just thermal equilibrium. 

By the area law \cite{bhd}, which includes $L_\xi A\cong0\Rightarrow H$ null, 
this also shows that {\em a black hole cannot change its angular momentum or 
charge without increasing its area}. 

On a null trapping horizon, one may take $\xi\cong\tau\cong l_+$, but the other 
null vector $l_-$ is non-unique, leading to non-uniqueness in $\omega$, used to 
define angular momentum here and for dynamical horizons by Ashtekar \& Krishnan 
\cite{AK1,AK2,AK3}. However, $\omega+\frac12Df$ is unique, an intrinsic normal 
fundamental form of a null hypersurface, therefore used to define angular 
momentum for isolated horizons by Ashtekar et al.\ 
\cite{ABF1,ABF2,AFK,ABD,ABL1,ABL2,Boo1,GJ1,DKSS}. On the other hand, the 
extrinsic normal fundamental form $\omega-\frac12Df$ is preserved 
\cite{bhd4,bhd5}: DEC $\Rightarrow L_+(\omega-\frac12Df)\cong0$. The 
non-uniqueness is therefore naturally fixed by 
\begin{equation}
Df\cong0.
\end{equation}
Recalling the definition (\ref{f}) of the normalization function $f$, this is a 
legitimate choice of gauge. Then all three normal fundamental forms coincide. 
Also $L_\xi\omega\cong0$, so that $L_\xi J\cong0$ assuming only that $\psi$ is 
a coordinate vector. Comparing with energy, NEC $\Rightarrow L_\xi M\cong0$ 
automatically on a null trapping horizon. 

Thus consideration of angular momentum resolves the ambiguity in taking the 
null limit. The general formalism for trapping horizons then applies in all 
cases, describing transitions between growing and non-growing phases of a black 
hole. This has largely recovered the notion of weakly isolated horizon due to 
Ashtekar et al.\ \cite{ABL1,ABL2,Boo1,GJ1}, except that the (allowable) scaling 
freedom in $\xi$ has not been fixed \cite{bhd4,bhd5}. 

\section{Type-changing horizons}
\begin{figure}
\includegraphics[height=5cm]{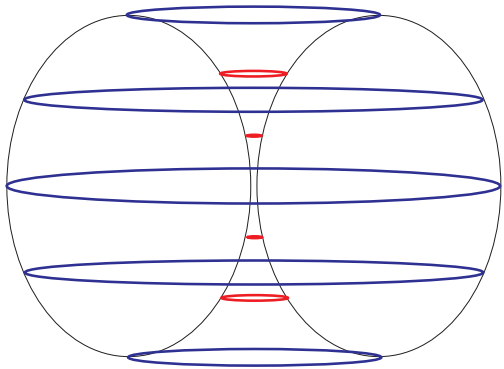}
\caption{Black-hole formation and evaporation. The vertical coordinate is 
loosely an advanced time, while the horizontal coordinates loosely indicate 
area, with marginal surfaces in the outer and inner trapping horizons depicted 
respectively as blue and red circles. In the lower half of the picture, the 
black hole is forming and the outer horizon is growing, while in the upper 
half, the black hole is evaporating and the outer horizon is shrinking. The 
inner horizon has the opposite properties. No attempt has been made to indicate 
timescales.} \label{evap} 
\end{figure}

While the above properties indicate that a future outer trapping horizon serves 
as a practical definition of black hole, it should be noted that a trapping 
horizon may change its type under evolution. For instance, in gravitational 
collapse, it is common for an inner horizon to form simultaneously with an 
outer horizon. In fact they join smoothly and are really two parts of a single 
horizon in a space-time sense, manifesting as distinct horizons in a spatial 
slicing (Fig.\ref{evap}). According to the basic laws, the transition occurs 
when the horizon is null, with the outer horizon achronal and the inner horizon 
causal. 

Booth et al. \cite{BBGV} constructed examples where an outer horizon becomes 
past-null, turning into a past-causal inner horizon, which can then turn into 
an outer horizon again, and so on. Also, numerical examples of Schnetter et al. 
\cite{SKB} show that a trapping horizon can be partly outer and partly inner on 
a given marginal surface, interpolating between regions where it is strictly 
inner or strictly outer. 

One application of such ideas is to evaporating black holes. Ingoing Hawking 
radiation tends to have negative energy density, $T_{++}<0$, violating NEC, and 
if it dominates over gravitational radiation in the sense that 
$T_{++}+\Theta_{++}<0$, the basic laws reverse: the outer horizon becomes 
causal and shrinks, while the inner horizon becomes spatial and grows. Thus it 
is possible that they simply reunite smoothly, closing off the region of 
trapped surfaces (Fig.\ref{evap}). The black hole has then evaporated. If no 
singularity ever formed, the space-time would have the global structure of 
Minkowski space-time and there would be no possibility of information loss and 
the associated information puzzle. Explicit examples of such space-times have 
been constructed \cite{02bh20}. 

\begin{figure}
\includegraphics[height=8cm]{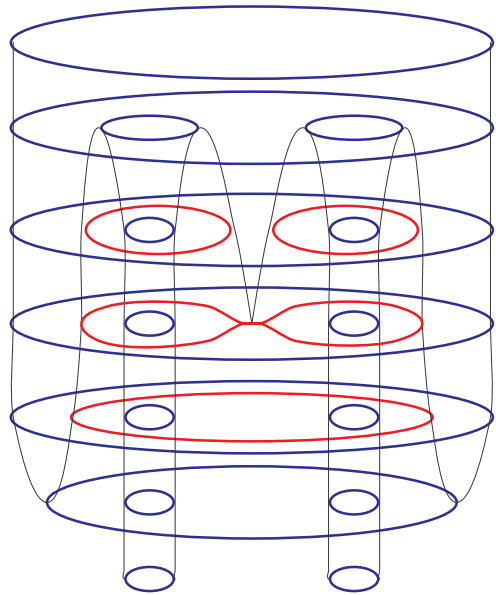}
\caption{Black-hole coalescence, with conventions as in Fig.\ref{evap}.
No attempt has been made to indicate relative motion.} 
\label{merger} 
\end{figure}

Another application is to the coalescence of binary black holes, of great 
interest as generators of strong gravitational waves. Typically a common outer 
horizon forms around the original two outer horizons. As above, it joins 
smoothly to an inner horizon, though the latter is not always tracked in 
simulations. It can be conjectured that the inner horizon later joins with the 
original outer horizons. For instance, the inner horizon may pinch off and 
change topology from one sphere to two spheres, which close in on the original 
outer horizons and merge smoothly with them, leaving the entire region inside 
the common outer horizon composed of trapped surfaces (Fig.\ref{merger}). This 
appears to be consistent with simulations \cite{SKB}, where the original outer 
horizons are slowly evolving and appear to be stable, while the inner horizon 
appears to be unstable and is rapidly evolving. More complex topologies are 
possible, and indeed there is numerical evidence that the original outer 
horizons can osculate and then intersect each other \cite{SPRTW}, allowing a 
self-intersecting inner horizon. It is also possible that singularities may 
intervene first, but this would be occurring in untrapped regions and so 
violate a version of strong cosmic censorship. 

According to the above picture, the entire set of horizons actually form a 
single trapping horizon in space-time, despite manifesting temporally as 
different and apparently unrelated horizons. The whole horizon can be smooth, 
though the spatial sections will have at least one point of non-smoothness 
where the topology changes. The irreducible mass $M$, being given by the area 
$A$, may be evaluated throughout the entire process, satisfying the energy 
conservation law. The separate areas $A_1$ and $A_2$ add to the common area $A$ 
at the topology change, $A=A_1+A_2$, so that, assuming NEC, 
\begin{equation}
A\ge A_1+A_2
\end{equation}
at any stage where either exists, where one follows the original outer horizons 
outwards in a space-time sense until the trapping horizon becomes a common 
outer horizon. The inequality holds also during inner stages, where one should 
note that the inner horizon is past-causal in an outward sense, so the area is 
still increasing outwards. This is an area-increase law for coalescing black 
holes, reminiscent of Hawking's famous theorem for event horizons \cite{Haw2}. 

It also implies 
\begin{equation}
M\ge\sqrt{M_1^2+M_2^2}.
\end{equation}
However, note that the total irreducible mass is actually reducible in such a 
process, indeed it necessarily decreases discontinuously at the change of 
topology, $M<M_1+M_2$. The greatest fractional loss of total irreducible mass, 
$(M_1+M_2-M)/(M_1+M_2)$, occurs in the equal-mass case and is 
$1-1/\sqrt2\approx29\%$, the same figure as obtained by the classical argument 
involving event horizons. In practice, the figure will be much less, since $M$ 
decreases only at the change of topology and is otherwise increasing outwards, 
particularly during the violent type-changing phase. 

Here one might prefer on physical grounds to use the effective energy $E$ 
instead of $M$ as a measure of available energy, which would require knowledge 
of angular momentum $J$. One may also evaluate $J$ throughout the whole 
process, but here there is an issue of whether it goes smoothly through the 
topology change, due to the more indirect construction and the numerically 
observed spin-flip phenomenon \cite{CLZKM}. This is of some interest by itself, 
as a potential way to relate initial and final angular momenta, and could be 
studied either analytically or by pushing simulations inside the region which 
is normally excised. 

\section{Concluding remarks}
It seems appropriate to conclude with some important recent results and issues. 
Firstly, there is the uniqueness of trapping horizons. Ashtekar \& Galloway 
\cite{AG} have shown that the structure of a given dynamical horizon is unique, 
in that there is only one way to foliate it by marginal surfaces. They also 
showed that such horizons are not so numerous as to foliate a space-time 
region, so that they will tend to interweave one another. 

Andersson, Mars \& Simon \cite{AMS1,AMS2} defined a strictly stable marginally 
outer trapped surface by a condition similar to the outer condition and showed 
that, given such a surface in one of a foliation of spatial hypersurfaces, such 
surfaces exist locally in the foliation. This shows that trapping horizons are 
not unique in a given space-time, indeed one is locally generated from a given 
marginal surface however the hypersurface is evolved. In practice, this is a 
useful quality since it means that a trapping horizon is likely to be found in 
numerical simulations, provided the foliation is chosen reasonably well. 

They also showed that, assuming NEC, the horizon is, on any one marginal 
surface, either spatial everywhere or null everywhere. Thus transitions between 
isolated and dynamical phases happen simultaneously with respect to the 
foliation. 

Andersson \& Metzger \cite{AM} have shown that, given a spatial hypersurface 
with an outer trapped inner boundary and an outer untrapped outer boundary, 
there exists a stable marginally outer trapped surface between, which is smooth 
and unique as the outermost such surface. This strengthens an earlier result 
which assumed piecewise smoothness \cite{KH}. Since this is the situation 
expected in numerical simulations of black holes, it guarantees the existence 
of an outermost trapping horizon with respect to the foliation. 

Another issue is that, in a spherically symmetric space-time such as Vaidya, it 
is possible to find trapped or outer trapped surfaces which lie partly outside 
the spherically symmetric trapping horizon, as shown numerically by Schnetter 
\& Krishnan \cite{SK} and analytically by Ben-Dov \cite{Ben}. The surfaces are 
roughly spherical and inside the horizon except for a long tentacle which 
extends through it. These are rather strange-looking surfaces, so one 
interpretation is that, in order to locate a black hole in a physically 
acceptable way, the type of trapping horizon needs to be refined further. For 
instance, it seems reasonable to require marginal or trapped surfaces to have 
positive Gaussian curvature, which would tend to exclude such tentacles. 
Alternatively or additionally, one might require any surface sufficiently close 
to a marginal surface and inside it to be a (non-strictly) trapped surface. 

This review has not addressed quantum issues directly, for which one may 
consult the review of Ashtekar \& Krishnan \cite{AK3}. One remarkable feature 
of the framework for isolated horizons is a derivation of black-hole entropy 
proportional to area $A$. Much less is known about dynamical situations. A very 
recent result is that, in spherical symmetry, a tunnelling method derives a 
local Hawking temperature $\kappa/2\pi$ precisely for future outer trapping 
horizons \cite{HCVNZ}. Thus this framework for dynamical black holes seems to 
be well adapted to the issues raised by Hawking radiation.

Research supported by the National Natural Science Foundation of China under 
grants 10375081, 10473007 and 10771140, by Shanghai Municipal Education 
Commission under grant 06DZ111, and by Shanghai Normal University under grant 
PL609.

\end{document}